\documentstyle[preprint,eqsecnum,aps]{revtex}
\begin{document}
\preprint{astro-ph/9501008}
\title{\Large\bf String Cosmology and Inflation.}
\author{Carlo Angelantonj$^1$, Luca Amendola$^{2,3}$,
Marco Litterio$^4$ and
Franco Occhionero$^2$}
\address{$^1$Dipartimento di Fisica, Universit\`a degli
Studi L'Aquila,
Via Vetoio, 67010 Coppito (AQ), Italy
\\
$^2$Osservatorio Astronomico di Roma, Viale del Parco Mellini 84,
I-00136
Roma, Italy
\\
$^3$NASA/Fermilab Astrophysics Center, Fermi National Accelerator
 Laboratory,
PO Box 500, Batavia IL 60510 - USA
\\
$^4$Istituto Astronomico, Universit\`a di Roma ``La Sapienza'',
Via Lancisi 29,
I-00161 Roma, Italy}
\date{\today}
\maketitle
\baselineskip 24pt
\begin{abstract}
Following a suggestion by Gasperini and Veneziano,  that
 String
Cosmology can be reconciled with Inflation and, hence, with the
Standard
Big Bang, we display an  analytical solution which possesses
four interesting properties: (1) it is non-singular; (2) it
distinguishes
the dynamics of the external scale factor, $a(t)$, from that of
 the internal
one, $b(t)$; (3) it exhibits a
non-monotonic behavior of $a(t)$; and (4) it
stabilizes both Newton's constant and $b(t)$ (the latter to a
finite,
non-vanishing value). The interest of the non-monotonic
evolution of $a(t)$
consists in the fact that it contains three phases of
accelerated expansion,
contraction and expansion before the final decelerated
expansion which
eventually becomes the Standard Big Bang. The total number
of e-folds of the
three accelerated eras can be calculated and tuned
to fit the requests of
observational astronomy.
\end{abstract}
\baselineskip 15pt
\section{Introduction.}
String Theory (ST) (see \cite{GSW} for a
review) is
at the moment the most attractive candidate
 for a unified description of the
basic constituents in nature and their
interactions. Despite tremendous progress
in our understanding of fundamental strings
in the past decade, we are still
very far from a single quantitative
prediction to be observed in experiments.
The main reason for this unsatisfactory
state of affairs is that the natural
scale in which string effects become
important (Planck scale) is much
smaller than the scale we can probe
in high energy scattering experiments.
However, as observational cosmology
provides a test of
fundamental physics, since the
scales of particle physics become
relevant as the universe grows
to its present size, the
most likely arena for a confrontation
between ST and experiments lies
in the cosmological inferences which
 may be
measurable today. To begin with, in general relativity,
singularities in curved space-times are
often unavoidable. In fact, the well-known
singularity theorems \cite{HE} prove
the existence of singularities under very general physical properties
of the matter
energy momentum tensor. For example, the Standard Big Bang (SBB)
scenario (see \cite{WB})  exhibits an initial
singularity at $t=0$. In ST,
there are several reasons to believe that singularities in target
space do not occur.
Heuristically, this belief is based on the fact that the
ST possesses a ``minimal length scale'' set by the
extension of the
string itself. One of the keys in understanding
the meaning of singularities and
 the minimal length scale in ST may be given by the so-called
duality symmetry \cite{TSD,AO}. Duality symmetry
is the most important
string symmetry from many points of view. Let us suppose we have a string
propagating in a target space ${\bf R}^{d-1} \times S^1$
where we have set the
radius of the compactified dimension equal to $R$
\cite{NSW}.
It is a well known fact that
every correlation function $A(1,\ldots ,N)$ can
 be written as a topological
expansion in the string coupling constant
\begin{equation}
A(1,\ldots ,N) =
\sum_{g=0}^{\infty} g_{st}^{2(g-1)} A_g (1,\ldots ,N)
\label{coorfunc}
\,,\end{equation}
where $A_g$ is the correlator at fixed genus. Duality
symmetry means that
$A(1,\ldots ,N)$ as a function of $R$ and $g_{st}$ is
invariant under the
replacement
\begin{equation}
R \to {\alpha ' \over R} \qquad \quad g_{st} \to
{\sqrt{\alpha  '} \over R}
g_{st} \label{cfds}
\,,\end{equation}
together with an interchange between the momentum and
the winding modes of the
external states. In other words, we are unable
to distinguish between small and
large $R$ provided we change the string
coupling properly. Since no string
scattering experiment is able to tell us
whether we are living in a universe
with size $R$ and string coupling constant
$g_{st}$ or in a universe with the
dual values, this defines in fact a minimal
measurable length at the self-dual
distance $\sqrt{\alpha '}$.
\par
The duality symmetry is not limited to flat
backgrounds; its existence was
shown \cite{VSFD} for
curved, time-dependent backgrounds, which is
of particular interest
in the context of cosmological singularities
\cite{PWUSC,PBBSC}. In addition to solve the
initial singularity problem,
a Theory
of Everything must also be able to explain
the low energy universe.
In particular in describing the history of
the whole space-time it must be able
to make contact with the SBB in the attempt
to describe properly the recent
evolution of the $4$-dimensional manifold.
The main difficulty we have to deal
with is that ST is defined on a $D$-dimensional
 manifold (with
$D=26$ or $10$ for the bosonic and
supersymmetric version respectively) while
we have experience of only three spatial plus
one temporal dimensions. Then the
theory must be able to describe the decoupling
of the external and internal
manifold. In particular we must require
${\cal M}^D = {\cal M}^4\times
{\cal K}^{D-4}$ where ${\cal M}^4$ is the
external 4-dimensional space-time and
${\cal K}^{D-4}$ is a $(D-4)$-dimensional
internal compact manifold
with typical physical dimension of
the order of Planck scale.
\par
Early attempts to find cosmologically
interesting string scenarios able to
eliminate the singularity in the early
history of the universe dates back to
the pioneering work of Alvarez, Leblanc,
Brandenberger and Vafa, Alvarez and
Osorio
\cite{PWUSC}, and were based on the hypothesis of target
space duality from thermodynamical considerations.
\par
More recently Gasperini and Veneziano
\cite{PBBSC} and Antoniadis, Rizos
and Tamvakis \cite{ART} proposed a
different dynamical approach to the solution
of the singularity problem, based on
Scale Factor Duality (SFD) and the
solution of the string effective equations.
\par
Both thermodynamical and dynamical
approaches proposed a suitable scenario in
which the evolution of the scale factor
is monotonic as time runs from
$-\infty$ to $+\infty$, while the Hubble
parameter is positive and
{\it bell shaped} as a function of time.
\par
In this paper, starting from dynamical
considerations, we show a different,
richer, non-singular scenario of string
cosmology able to make contact with the
SBB for the late evolution of the universe.
In particular,
after introducing external and internal
scale factors, $a(t)$ and $b(t)$,
we find that the
evolution of $a(t)$ is {\it not} monotonic;
a contraction phase
is present
in the early history of the universe,
while the effective 4-dimensional
gravitational coupling naturally
converges to a constant value corresponding to
the present value of Newton's constant.
As during the evolution of the universe
there are
different phases of accelerated dynamics,
the scenario presented in this
paper, in addition to solving the problems
of the singularity and of the
constancy of the
fundamental constants, offers a natural
framework in which to
accomodate inflation
(which is the solution of the other well
known problems of the SBB, i.e.
horizon, flatness and structure formation)
(see, however, Ref. \cite{BS}).
\par
This paper is organized as follows:
in Sect. II we discuss the low energy string
effective action and the
SFD symmetry. Sect. III is devoted to a
general and qualitative description
of solutions of the string field equations.
In Sect. IV we present the
general solution to the string cosmology
equations which interpolates smoothly
the dual evolution of the `pre-big-bang'
and `post-big-bang' phases,
where here by big bang (in low case)
we mean the epoch of transition between
the two dual phases. In Sect.
V we summarize the main conclusions.
The Appendixes are devoted to the
proof of the non-singular behavior of
the solution reported in Sect. IV,
and to the representation of the same
solutions in the Einstein frame,
respectively.
\section{Low Energy String Effective Action and Scale Factor Duality.}
\par
Let us consider the propagation of a
bosonic string in the presence of a
background consisting of a
$D$-dimensional metric $g_{\mu\nu}$ ($\mu ,\nu = 0,1
\ldots ,D-1$)
and a dilaton $\Phi$. It is described by the two-dimensional
$\sigma$-model
\begin{equation}
S_{\sigma}  = \int d^2 x \sqrt{h}
\left[ h^{ab} \partial_a X^\mu \partial_b
X^\nu g_{\mu\nu} (X^\rho ) +
\alpha ' R^{(2)} \Phi (X^\rho ) \right]
\label{tdsm}
\,,\end{equation}
where $h_{ab}$ is the world-sheet
metric tensor and $R^{(2)}$ is the Ricci
scalar constructed with $h_{ab}$.
The requirement of conformal invariance of
$S_\sigma$ (i.e. the vanishing of
the $\beta$-functions) leads naturally to the
determination of the massless modes'
dynamics. In particular we have the
following tree level effective action
for the background fields \cite{LESEA}
\begin{equation}
S_{{\rm eff}} = -{1 \over 2\kappa^2} \int d^D x \sqrt{-g}
e^{-\Phi} \left( R +
\partial_\mu \Phi \partial^\mu \Phi + c \right)
\label{lesea}
\,,\end{equation}
a multidimensional Brans-Dicke (BD) theory.
The cosmological constant represents the central
charge deficit of the theory,
$c = - 2 (D_{{\rm eff}} -D_{{\rm crit}})/3 \alpha '$
depending on details
of particular ST ($D_{{\rm eff}} =D$, $D_{{\rm crit}}=26$
in the bosonic
version,
$D_{{\rm eff}} ={\textstyle {3\over 2}}D$, $D_{{\rm crit}}=15$
in the supersymmetric version). The effective
action (\ref{lesea}) leads to the
following equations of motion
\begin{mathletters}
\label{lesfe}
\begin{eqnarray}
0 &=& (\partial\Phi )^2 - 2 \Box \Phi - R - c\,, \label{lesfe a}
\\
0 &=& \left( R_{\mu\nu} + \nabla_\mu \nabla_\nu \Phi \right) e^{-\Phi}
\label{lesfe b}
\end{eqnarray}
\end{mathletters}
($\nabla_\mu$ is the covariant derivative
and $\Box= \nabla_\mu \nabla^\mu$
is the
$D$-dimensional d'Alambertian).  Now, it
is known \cite{VSFD,SFD} that if the metric
and dilaton fields do not depend on
the coordinate $x^i$, the field equations
(\ref{lesfe}) are invariant under the
SFD transformation
\begin{mathletters}
\label{sfdt}
\begin{eqnarray}
g_{ii} & \to & \widetilde{g} _{ii} = g_{ii}^{-1} \,,\label{sfdt a}
\\
\Phi & \to & \widetilde{\Phi} = \Phi - \ln | g_{ii} | \label{sfdt b}
\,,\end{eqnarray}
\end{mathletters}
The non-trivial duality tranformation
behavior of the dilaton field implies
that the coordinate-dependent string
coupling constant is trasformed like
$g_{st}^{2} (x) =e^{\Phi} \to g_{st}^{2} (x) g_{ii}^{-1}$.
 This change of the
string coupling constant agrees with
the transformation of $g_{st}^{2}$ in the
static case (equation (\ref{cfds}))
when one considers the genus expansion of
the string partition function \cite{AO}.
\par
This transformation is just a particular
case of a more general global $O(d,d)$
covariance of the theory \cite{ODDMVI,ODDGV,ODDMVII}:
$O(d,d)$ covariance means that,
if the theory is independent of $d$ spatial
coordinates, the dilaton tranforms as
\begin{equation}
\Phi \to \Phi - \ln | \det g_{ij} |
\label{odddt}
\,,\end{equation}
and the components of the metric and of
the antisymmetric\footnote{The third
massless mode of the bosonic string.}
tensors mix according to
\begin{equation}
M \to \Omega^{{\rm T}} M \Omega
\label{oddma}
\,,\end{equation}
where $\Omega \in O(d,d)$ and
\begin{equation}
M = \left( \matrix{ G^{-1} & -G^{-1} B \cr BG^{-1} &
G- BG^{-1} B \cr} \right)
\label{mmt}
\end{equation}
($G\equiv g_{ij}$ and $B\equiv B_{ij}=-B_{ji}$
are matrix representations of
the $d\times d$ spatial part of the metric
and antisymmetric tensor, in the
basis where the $O(d,d)$ metric is off-diagonal.)
\par
In reference \cite{ODDGV} it was shown that
$O(d,d)$ covariance holds even if the
equations (\ref{lesfe}) are supplemented
by a phenomenological source term
corresponding to bulk string matter.
\par
The importance of SFD in the context of string cosmology
is that, when combined with time reversal (the most obvious
symmetry of the theory), it
allows us to associate at every phase of
`post-big-bang' evolution (for $t_c
<t<+\infty$) a dual phase, called `pre-big-bang' \cite{PBBSC},
($-\infty <t <
t_c$) with a completely different dynamics
for the fields. In fact SFD is not a
simple reparametrization of the fields,
nor are its implications
trivial. For example, if we start with a scale
factor $a(t)$ that is expanding, the dual scenario $a (t) \to a^{-1} (t)$
describes a contracting universe.
When combined with time reversal, SFD maps, for
example, a background with decreasing curvature to the dual
one characterized by
a curvature that is increasing.
\par
Finally it is important to stress the necessity of the
presence of the dilaton for SFD to
be a symmetry of the low energy string effective action.
In fact for $\Phi
\equiv 0$ the action (\ref{lesea}) reduces to conventional general
relativity, and we are left simply with the symmetry
of time reversal. Only the
existence of the dilaton field and its non
trivial behavior under SFD (equation
(\ref{sfdt})) allows us to realize the transformation $a \to a^{-1}$.
\section{General Cosmological Solutions.}
To make contact with observational cosmology,
we take the $D$-dimensional
space-time as a direct product of the external
 pseudo-Riemannian manifold
${\cal M}_4$ with an $n(=D-4)$-dimensional
compact Riemannian manifold
${\cal K}_n$. We
take for ${\cal M}_4$ a flat
Friedman-Robertson-Walker space-time with scale factor $a(t)$, and
we assume that the dilaton
field $\Phi$ and the radius $b$ of the
internal space (which we take to be an
$n$-torus) depend only on the temporal coordinate:
\begin{eqnarray}
g_{\mu\nu} &=& {\rm diag} (1 ,-a^2 (t) \delta_{ij} , -b^2 (t) \delta_{ab} )
\,,\\
\Phi &=& \Phi (t)
\end{eqnarray}
($i,j =1,2,3;\ a,b=1,\ldots ,n=D-4$). With
these choices for the fields and
adding a source term representing a
primordial string gas with energy momentum
tensor
\begin{equation}
T_{\mu}^{\nu} = {\rm diag} (\rho (t)\, , \, - p(t) \delta_{i}^{j}\, , \,
-q(t) \delta_{a}^{b}
)\, ,
\label{emt}
\end{equation}
the field equations (\ref{lesfe}) (with $c =0$) which now change
in\footnote{These three equations are not independent. The third
one, in fact, can be obtained by a combination
of the gradient of the first
and the second.}
\begin{mathletters}
\label{lesfem}
\begin{eqnarray}
(\partial\Phi )^2 - 2 \Box \Phi - R &=& 0 \label{lesfem a}\,,
\\
 \left( R_{\mu}^{\nu} + \nabla_\mu \nabla^\nu \Phi \right) e^{-\Phi}
&=&  \kappa^2  T_{\mu}^{\nu} \,,
\label{lesfem b}
\\
\nabla^\mu T_{\mu}^{\nu} &=& 0\,,
\label{lesfem c}
\end{eqnarray}
\end{mathletters}
can be written as
\begin{mathletters}
\label{rfe}
\begin{eqnarray}
-2 \ddot{\bar\Phi} + \dot{\bar\Phi} ^2 + 3 H^2 + n F^2 &=& 0 \,,
\label{rfe a}
\\
\dot{\bar\Phi} ^2 - 3 H^2 - n F^2 &=& \kappa^2 e^{\bar\Phi} \bar\rho\,,
\label{rfe b}
\\
\dot H - H \dot{\bar\Phi} &=& \kappa^2 e^{\bar\Phi} \bar p\,,
\label{rfe c}
\\
\dot F - F \dot{\bar\Phi} &=& \kappa^2 e^{\bar\Phi} \bar q\,,
\label{rfe d}
\\
\dot{\bar\rho} + 3 H \bar p + n F \bar q &=& 0\, .
\label{rfe e}
\end{eqnarray}
\end{mathletters}
Here we have introduced the Hubble parameters
$H=\dot a /a$ and $F=\dot b /b$
for the external and internal space respectively
and we have
denoted with barred symbols
the $O(D-1,D-1)$-invariant
expressions for the dilaton and the matter
energy density\footnote{The
behavior of the latter under SFD consists
in \cite{ODDGV} $\bar\rho \to
\bar \rho$, while for the pressures
$p/\rho \to - p /\rho$ and
$q/\rho \to - q /\rho$.}
\begin{eqnarray}
\bar\Phi &=& \Phi - 3 \ln a - n \ln b\,,
\\
\bar\rho &=& \rho a^3 b^n
\end{eqnarray}
(we also introduce $\bar p = p a^3 b^n$ and
$\bar q = q a^3 b^n$.)
To solve the system (\ref{rfe}) we must
introduce an equation of
state for the source term of the form
\begin{equation}
p = \gamma \rho \, , \qquad q=\lambda \rho
\label{ste}
\,,\end{equation}
with, at the moment,  $\gamma$ and $\lambda$ arbitrary functions of
time.
\par
Introducing a coordinate time $\xi$ defined by
\begin{equation}
d\xi  =\bar\rho \ell d t
\,,\end{equation}
($\ell$ is an arbitrary constant of dimension of length)
the forementioned
equations can be integrated \cite{PBBSC,ODDMVII}
to obtain ($2 \kappa^2 =1$)
\begin{mathletters}
\label{equat}
\begin{eqnarray}
\bar\Phi - \Phi_0 &=& - 2 \int^{\xi}_{0}
{\xi - \xi_0 \over \Delta (\xi )}\, d \xi\,,
\label{equat a}
\\
\bar\rho  &=& \case{1}/{4}\ell^{-2} \Delta (\xi ) e^{\bar\Phi}\,,
\label{equat b}
\\
H &=& \case{1}/{2} \ell^{-1} (\alpha_H + \Gamma ) e^{\bar\Phi}\,,
\label{equat c}
\\
F &=& \case{1}/{2} \ell^{-1} (\alpha_F + \Lambda ) e^{\bar\Phi}\,,
\label{equat d}
\end{eqnarray}
\end{mathletters}
where
\begin{equation}
\Gamma = \int^{\xi}_{0} \gamma\, d \xi\,,\qquad
\Lambda = \int^{\xi}_{0} \lambda \, d \xi\,,
\end{equation}
\begin{equation}
\Delta (\xi ) = 4 \beta + (\xi + \xi_0 )^2 -
6 \alpha_H \Gamma - 3 \Gamma^2  - 2
n \alpha_F \Lambda - n \Lambda^2\, ,
\label{dxi}
\end{equation}
$\Phi_0 ,\xi_0 , \alpha_H , \alpha_F$ are arbitrary
constants and
\begin{equation}
\beta = -
\case{1}/{4} ( 3 \alpha_{H}^{2} + n \alpha_{F}^{2} )
\label{bet}
\,,\end{equation}
is negative.
\par
{}From (\ref{equat b}), to have a positive
definite energy
density we must require $\Delta (\xi ) >0$;
also the zeroes of $\Delta (\xi )$ correspond
to singularities for the fields.
\par
Let us consider the simplest
 case $\gamma =\hat{\gamma}$ and $\lambda =\hat{\lambda}$,
where $\hat{\gamma}$ and $\hat{\lambda}$ are constants.
The general solution of
(\ref{equat}) \cite{IDFI,DPSC} in a form convenient
to our discussion reads
\begin{mathletters}
\label{gss}
\begin{eqnarray}
a(\xi ) &=& a_0 \left| (\xi -\xi_+ )(\xi - \xi_-
)\right|^{\hat{\gamma}
/\epsilon} \left| {\xi - \xi_+ \over \xi -\xi_-}
\right|^{\sigma_H}
\,,\label{gss a}
\\
b(\xi ) &=& b_0 \left| (\xi -\xi_+ )(\xi - \xi_-
)\right|^{\hat{\lambda}
/\epsilon} \left| {\xi - \xi_+ \over \xi -\xi_-}
\right|^{\sigma_F}
\,,\label{gss b}
\\
H(\xi ) &=& \case{1}/{2} \ell^{-1} e^{\Phi_0}
(\alpha_H + \hat{\gamma} \xi )
\left| (\xi -\xi_+ )(\xi - \xi_- )\right|^{-1/\epsilon}
\left|
{\xi - \xi_+ \over \xi -\xi_-} \right|^{-\sigma_0}
\,,\label{gss c}\\
F(\xi ) &=& \case{1}/{2} \ell^{-1} e^{\Phi_0}
(\alpha_F + \hat{\lambda} \xi )
\left| (\xi -\xi_+ )(\xi - \xi_- )\right|^{-1/\epsilon}
\left|
{\xi - \xi_+ \over \xi -\xi_-} \right|^{-\sigma_0}
\,,\label{gss d}\\
e^{\Phi (\xi )} &=& a_{0}^{3} b_{0}^{n} e^{\Phi_0}
\left| (\xi -\xi_+ )(\xi - \xi_- )\right|^{-(1-3\hat{\gamma}
-n\hat{\lambda})/\epsilon} \left|
{\xi - \xi_+ \over \xi -\xi_-} \right|^{-\sigma_0+3\sigma_H
+n \sigma_F}
\,,\label{gss e} \\
\rho (\xi ) e^{\Phi (\xi )} &=& \case{1}/{4} \epsilon
\ell^{-2} e^{2\Phi_0}
{\rm sign} [\Delta (\xi )]
\left| (\xi -\xi_+ )(\xi - \xi_- )\right|^{(\epsilon -2)/\epsilon}
\left|
{\xi - \xi_+ \over \xi -\xi_-} \right|^{-2\sigma_0}
\,,\label{gss f}
\end{eqnarray}
\end{mathletters}
where
\begin{equation}
\xi_\pm = {1 \over \epsilon} \left\{  3
\alpha_H \hat{\gamma} + n\alpha_F \hat{\lambda} -
\xi_0 \nonumber
 \pm \left[ \left( \xi_0 - 3\alpha_H \hat{\gamma} - n \alpha_F
\hat{\lambda} \right)^2 - \epsilon
\left( \xi_0 -3 \alpha_{H}^{2} - n \alpha_{F}^{2} \right) \right]^{1/2}
\right\}
\,,\end{equation}
are the two {\it real} zeroes of $\Delta (\xi )$
and
\begin{eqnarray}
\epsilon &=& 1 - 3\hat{\gamma}^2 - n \hat{\lambda}^2 \,,\label{epsi}
\\
\sigma_0 &=& {\xi_+ + \xi_- - 2 \xi_0 /\epsilon \over \xi_+ - \xi_-}\,,
\nonumber\\
\sigma_H &=& {\xi_+ + \xi_- + 2 \alpha_H /\epsilon \over \xi_+ - \xi_-}
\,,\nonumber \\
\sigma_F &=& {\xi_+ + \xi_- + 2 \alpha_F /\epsilon \over \xi_+ - \xi_-}
\,.\nonumber
\end{eqnarray}
Other solutions of the system (\ref{rfe}) are also obtained from
(\ref{gss})  through SFD.
\par
The internal and external Hubble parameters
have two singularities,  at $\xi=\xi_+$ and at $\xi=\xi_-$. For
$\epsilon >0$ (necessary condition to have a positive
 energy density today) the
range $(\xi_- ,\xi_+ )$ is not physical because here $\rho$
becomes negative. The evolution of the scale
factors depends strongly on the
relative sign of $\alpha_H$ and $\hat\gamma$, and of $\alpha_F$ and
$\hat\lambda$; but in any case singularities are
always present in the
curvature, contrary to what stressed in \cite{BUS} in the context of
Brans-Dicke
theory. In Fig.\ \ref{fig1} we report a qualitative
representation of a generic scale factor for $\alpha_i / \omega_i >0$ and
$\alpha_i / \omega_i <0$, where
$\{\alpha_i \} = (\alpha_H , \alpha_F )$ and
$\{\omega_i \} = (\hat\gamma , \hat\lambda )$.

\section{Non-Singular Solutions.}

For the solutions found in Sect. III the growth of curvature,
of the effective coupling $e^\Phi$ and of the effective energy
density $\rho e^\Phi$, are unbounded, which is
inacceptable in the light
of the discussion of
Sect. I and to phenomenological constraints on the graviton spectrum
discussed in \cite{PBBSC}. The problem is then to
find a smooth transition from the pre-big-bang phase to the
post-big-bang one. For this purpose we exploit a suggestion of Gasperini
and Veneziano \cite{PBBSC}.
\par
In the vicinity of the Planck scale the low energy
string effective action
(\ref{lesea}) does not apply; we expect
some modifications. To preserve the
symmetry under SFD we must require that
these corrections are themselves
invariant. Following \cite{PBBSC} we
introduce a self-dual dilaton potential
$V (\bar\Phi ) = -V_0 e^{2\bar\Phi}$.
The new field equations can be still reduced
to the form (\ref{equat}), but with
\begin{equation}
\beta = \ell^2 V_0 - \case{1}/{4} (3
\alpha_{H}^{2} + n \alpha_{F}^{2} )
\label{bett}
\,,\end{equation}
which, unlike (\ref{bet}), is no longer necessarily negative.
For $\epsilon >0$ (positivity of the source
energy density) it is then possible to choose $V_0 >0$ and large
such that $\Delta (\xi )$ does
not have zeroes in the real field (a proof is given in Appendix A).
This implies that
there are no  singularities in the curvature,
nor in the effective coupling,
nor in the effective
energy density.
\par
To solve the system (\ref{equat}) we must assign
the equation of state of the
source term. Being interested in solutions which
describe a smooth transition
from the pre-big-bang era ($\xi <0$) to the dual
post-big-bang ($\xi >0$) it
is  worth looking for self-dual solutions, and in
order to obtain them,
as $\xi$
goes through zero, the external and internal
pressures must change sign,
\cite{PBBSC}.
So $\gamma (\xi )$ and $\lambda (\xi )$ must
be odd functions of $\xi$. From
phenomenological considerations we must also
impose the constraint of
stationariety of the functions $\gamma$
and $\lambda$ for large values of
$|\xi|$. Then, if we introduce the vector of
 pressure $\{ p_i \} = (p,q)$,
we must require
\begin{eqnarray}
{p_i \over \rho} &\to & - \omega_i \,,\qquad {\rm for}\ \
\xi <-\bar\xi\,,\nonumber\\
{p_i \over \rho} &\to & + \omega_i \,,\qquad
{\rm for}\ \ \xi >+\bar\xi\,,\nonumber
\end{eqnarray}
($\omega_i$ and $\bar\xi$ constants) with a smooth
transition between the two
phases. It is evident that these properties are
shared by a vast class of
functions. For example, any of the choices
\begin{equation}
{f(\xi ) \over \sqrt{f^2 (\xi ) + \hat\xi ^2}}\, , \quad \tanh (f(\xi ))\, ,
\quad {2 \over \pi} \arctan (f(\xi ))\,, \quad \ldots
\end{equation}
with $f(\xi )$ arbitrary, smooth, odd and asymptotically monotonic
 function of $\xi$, would do.
But, since, from a qualitative point of
view  \cite{TESI}, the
result is independent of the form of
the functions $\gamma\, ,\lambda$ for
fixed $f$, for mathematical simplicity,
we choose (as in \cite{PBBSC})
\begin{equation} \gamma (\xi ) = {\hat\gamma \xi
\over \sqrt{\xi^2 +\xi_{1}^{2}}} \, ,
\qquad \lambda (\xi ) = {\hat\lambda \xi
\over \sqrt{\xi^2 +\xi_{2}^{2}}}
\,,\end{equation}
with $\hat\gamma , \hat\lambda ,
\xi_1 , \xi_2$ constants.
\par
Equations (\ref{equat}) can be solved
analytically if $\xi_0 =0$ and $\xi_1
=\xi_2$. In place of the singular
solutions (\ref{gss}) we obtain
now \cite{TESI}
\begin{mathletters}
\label{unsol}
\begin{eqnarray}
H (\xi ) &=& \case{1}/{2} \ell^{-1} e^{\Phi_0} \left( \alpha_H +
 \hat\gamma
\sqrt{\xi^2 + \xi_{1}^{2}} \right) (\Delta (\xi ))^{-1/\epsilon}
\exp \left\{ -
{2\zeta \over \epsilon \sqrt{\chi}} \arctan
\left( {\epsilon \sqrt{\xi^2
+\xi_{1}^{2}} - \zeta \over \sqrt{\chi}} \right) \right\}
\,,\label{unsol a}\\
F (\xi ) &=& \case{1}/{2} \ell^{-1} e^{\Phi_0}
\left( \alpha_F + \hat\lambda
\sqrt{\xi^2 + \xi_{1}^{2}} \right) (\Delta (\xi ))^{-1/\epsilon} \exp
\left\{ -
{2\zeta \over \epsilon \sqrt{\chi}} \arctan
\left( {\epsilon \sqrt{\xi^2
+\xi_{1}^{2}} - \zeta \over \sqrt{\chi}} \right)
 \right\}
\,,\label{unsol b}
\\
e^{\bar\Phi (\xi)} &=& e^{\Phi_0} (\Delta (\xi ))^{-1/\epsilon}
\exp \left\{ -
{2\zeta \over \epsilon \sqrt{\chi}} \arctan \left( {\epsilon
\sqrt{\xi^2
+\xi_{1}^{2}} - \zeta \over \sqrt{\chi}} \right) \right\}
\,,\label{unsol c}\\
\bar\rho (\xi ) &=&
\case{1}/{4} \ell^{-2} e^{\Phi_0} (\Delta (\xi ))^{(\epsilon
-1)/\epsilon} \exp \left\{ - {2\zeta \over \epsilon \sqrt{\chi}}
 \arctan \left(
{\epsilon \sqrt{\xi^2 +\xi_{1}^{2}} - \zeta \over \sqrt{\chi}}
\right) \right\}
\,,\label{unsol d}
\end{eqnarray}
\end{mathletters}
where
\begin{eqnarray}
\zeta &=& 3\alpha_H \hat\gamma + n \alpha_F \hat\lambda
\,,\\
\chi &=& (4 \beta - \epsilon \xi_{1}^{2}) \epsilon - \zeta^2
\,,\end{eqnarray}
and $\epsilon$ has been defined in (\ref{epsi}).
For the scale factors, as we are unable to give analytical
expressions, we have proceeded to numerical integrations.
In any case a qualitative information can be
easily obtained from the
relative Hubble parameters. In particular, if we define
$\{ H_i (\xi ) \} = \left( H(\xi ),F(\xi ) \right)$,
it is easy to see that if ${\rm sign} [\alpha_i \omega_i ] =+1$
then the Hubble
parameter $H_i$ never changes sign, while if
${\rm sign} [\alpha_i \omega_i ]=-1$
then
\begin{equation}
{\rm sign} \left[ H_i \left( |\xi | < \sqrt{ \left( {\alpha_i \over \omega_i}
\right)^2 - \xi_{1}^{2} }\right) \right] = -
{\rm sign} \left[ H_i \left( |\xi | > \sqrt{ \left( {\alpha_i \over \omega_i}
\right)^2 - \xi_{1}^{2} }\right) \right]
\,.\end{equation}
Therefore, if for large $|\xi |$, $H_i$ is positive (corresponding
to a background that
today is expanding), for $\xi$ near zero we have a transition {\it
expansion-contraction-expansion} as depicted in Fig.\ \ref{fig2}.
\par
These solutions generalize in a non-trivial way the ones obtained in
\cite{PBBSC}. The latter, in fact, can be easily recovered if we set
$\alpha_H =
\alpha_F =0$, $\hat\gamma =-\hat\lambda =1/(3+n)$ and
$4\beta =\xi_{1}^{2}$.
\par
The presence of a contraction phase, very suggestive in itself,
has interesting consequences on the present structure of the
observable universe. It was recently stressed in \cite{IDFI}
that every kind of
accelerated evolution, whether during expansion or contraction,
naturally solves the
kinematical problems of the SBB. Moreover, when expressed
in terms of conformal
time, the contraints for successful inflation are the
same for expansion and
contraction. Then our solution offers a natural
scenario capable of solving the
flatness, horizon and structure formation problems of
SBB, without introducing
an {\it ad hoc} inflaton.
As for the number of e-folds of accelerated contraction, we have
\begin{equation}
{\cal N} \equiv \int_{t_{{\rm in}}}^{t_{{\rm fin}}} H(t)\, dt
=\int_{\xi_{{\rm in}}}^{0}
H(\xi ) {dt \over d\xi} \, d\xi = \int_{\xi_{{\rm in}}}^{0} W(\xi )\, d\xi
\,,\end{equation}
with $\xi_{{\rm in}} =-\sqrt{(\alpha_H /\hat\gamma )^2 -\xi_{1}^{2}}$ the
negative zero of $H(\xi )$ and the integrand $W(\xi )
\equiv 2 (\alpha_H +\Gamma )/\Delta (\xi )$.
Although the latter integral cannot be
performed analytically, it can be seen that $W(\xi )$
has the same shape of
$H(\xi )$ and, for the period of accelerated contraction, a very good
fitting is a linear interpolation (we have a correlation
coefficient $R=0.99\ldots$). Then we get for the linear
fitting function $\widehat W$ in the range
$\xi \in [\xi_{{\rm in}} , 0]$
\begin{equation}
\widehat W (\xi ) = - W(0) \left( {\xi \over \xi_{{\rm in}}} -1
\right)
\,,\end{equation}
and
\begin{equation}
{\cal N} = \case{1}{2} W(0) \xi_{{\rm in}}
\,,\end{equation}
where
\begin{equation}
W(0) \equiv W(\xi =0)=
 {\alpha_H + \hat\gamma |\xi_1 | \over 4 \ell^2 V_0
- (3 \hat \gamma ^2 + n \hat\lambda ^2 ) \xi_{1}^{2} - 2 ( 3 \alpha_H
\hat\gamma + n \alpha_F \hat\lambda ) |\xi _1 | - 3 \alpha_{H}^{2}
- n \alpha_{F}^{2}}\,.
\end{equation}
We can choose $\alpha_H$ and $\alpha_F$ to tune ${\cal N}$ to any desired
value.
But the contraction phase is not the only period
of accelerated evolution of the external space. For
${\rm sign} [\alpha_H
\hat\gamma ] =-1$ three different phases of accelerated evolution
(see Fig.\ \ref{fig3}) follow each other during
 the early history of the universe. Then,
in addition to solving naturally
the kinematical problems typical of the SBB, from the
point of view of structure formation, our scenario gives rise  to
a sort of  multiple
inflation (but with a single field)
capable of
breaking the scale invariance of the fluctuation power spectrum,
a possible solution to the problem of large scale power in
galaxy distribution \cite{GM}.
\par
Furthemore, in order to make contact with
observational cosmology, in addition to explaining why the
universe is flat, the entropy is so high, etc., we must
also be able
to explain
why the fundamental constants are effectively constants.
In fact in
string cosmology the gravitational coupling is dynamical.
When reducing the theory from $D$
to four dimensions, we get that the 4-dimensional
``Newton's constant'' is
proportional to the inverse of the volume of the internal space times the
inverse of the coupling between the scalar field and the Ricci scalar
($G_N
\sim b^{-n} e^{\Phi}$): in general, the latter
expression is hardly constant,
while Newton's constant must be ``constant'' at least from nucleosynthesis
onward \cite{LVNC}.
Really this is the most difficult problem in
multi-dimensional and scalar-tensor
theories.
\par
If we look at the asymptotic behavior of solutions
(\ref{unsol}), then we can
realize that our scenario has just this additional bonus.
It is in fact easy
to see that for very large $\xi$ we have
\begin{mathletters}
\label{asyms}
\begin{eqnarray}
a (\xi ) & \sim & \xi^{2\hat\gamma /\epsilon}
\,,\label{asyms a}
\\
b (\xi ) & \sim & \xi^{2\hat\lambda /\epsilon}
\,,\label{asyms b}
\\
e^{\Phi (\xi )} & \sim & \xi^{2(3\hat\gamma + n
\hat\lambda -1)/\epsilon}
\,,\label{asyms c}
\end{eqnarray}
\end{mathletters}
and
\begin{equation}
G_N \sim b^{-n} e^{\Phi} \sim \xi^{2 (1-3\hat\gamma )/\epsilon}
\,. \label{asymg}
\end{equation}
Then for a universe with radiation in the external space
$\hat\gamma
=\case{1}/{3}$, $G_N$ becomes asymptotically constant (see
Fig.\ \ref{fig4}). What is really surprising is that
Newton's constant stabilizes {\it indipendently} of
the dynamics of the internal
space, which asymptotically
%
can expand (for $\hat\lambda >0$), contract (for
$\hat\lambda <0$) or approach a constant value (for
$\hat\lambda =0$; the most attractive possibility).
For what concerns the asymptotic dynamics of the external
 space, the choice
$\hat\gamma =\case{1}/{3}$ naturally leads to the typical
behavior of Radiation
Dominated model:
$a(t) \sim t^{1/2}$.\par
The same asymptotic scenario is shared also by the
singular solutions
(\ref{gss}), because the dilaton potential strongly
modifies the field's dynamics only around $|\xi |\sim 0$,
becoming rapidly
uninfluential as $\xi$ grows.
\section{Conclusions.}
The combination of
Einstein's general theory of relativity and of the
Copernican principle naturally leads
to the formulation of the SBB.
%
Although
in the last decades the SBB has been strongly
confirmed by astronomical observations,  it still presents
some ``conceptual'' difficulties: the existence of an
initial singularity and
the well known kinematical problems (horizon, flatness,
structure formation).
It is a common belief that these problems can be
solved when a consistent
quantum theory of gravitation will be formulated.
Today ST seems the
most plausible attempt to quantize gravity, and
then it is very tempting to
study its implications on the early evolution
of the universe. Foremost in
supporting the belief that string cosmology
can solve the singularity problem, is
SFD symmetry, one of the most important symmetries of ST. It means
that if $a(t)$ solves the string equations
then also $a^{-1} (t)$ is
a solution of the same dynamical system, thus
introducing a minimal length
scale. Furthermore ST modifies the commonly
accepted lore: the present
decelerated expansion is preceded by a dual
phase in which the evolution is
accelerated. The smooth passage between the
two asymptotic phases may be
realized by a period of accelerated contraction.
Because accelerated
contraction is as efficient as accelerated
expansion to solve the kinematical
problems of the SBB, the ST scenario presents   multiple
episodes of inflation. The
last difficulty we are able to cope with is to
justify the constancy of the
4-dimensional Newton's constant without requiring the
introduction of a mass term
for the dilaton or the formulation of a least
coupling principle \cite{DP}. In
fact the low energy string effective action
is a multi-dimensional scalar tensor
theory of gravity, presenting then a dynamical
gravitational coupling.
Astronomical observations imply however that
$\dot G _N / G_N \alt 10^{-11}
{\rm sec}^{-1}$ \cite{LVNC}: a realistic model
must describe the spontaneous
stabilization of the Newton's constant, which we obtain.
\par
Obviously the scenario presented here is basically a toy model,
which needs more
theoretical support. For example we should
justify the form of the dilaton
potential and the equation of state for
string sources in curved backgrounds.
Nevertheless we want to stress that some
of the positive and new results
presented in this paper (presence of a
primordial contraction phase,
stabilization of the fundamental constants,
convergence towards the SBB) are
common both to non-singular and to more
conventional singular scenarios,
representing then a element in favour of string cosmology.
\par
The development of the subject and the
deeper study of the astrophysical
implications of an early dynamical
gravitational coupling and contraction
phase, in addition to the spontaneous
compactification of
the internal dimensions, may represent
a solid benchmark
to test the theoretical predictions of ST.
\par
\vspace{.3in}
\centerline{\bf Acknowledgment}
\vspace{.1in}
C.A. is very grateful to A. Sagnotti and M. Gasperini for
stimulating discussions and comments.
The work of L.A. at Fermilab was
supported by DOE and NASA under grant NAGW-2381. L.A.  also
acknowledges   CNR (Italy) for financial support.
\newpage

{\it Note added in proof.}

While completing this work we became aware of some recent
literature which we believe relevant to the problem.
J.J. Levin and K. Freese [Phys. Rev. D {\bf 47}, 4282 (1993)],
discuss an inflationary   scalar-tensor theory of gravity
which, like the one we presented in the text,
does not require
 a scalar field potential.
E. J. Copeland, A. Lahiri \& D. Wands
[preprint SUSX-TH-94/3-7, and preprint SUSSEX-AST-94/10-1],
describe cosmological models containing the
antisymmetric rank-2 tensor field, $B_{\mu\nu}$,
which show a singular dynamics similar to the one in Sec. III.
Finally, J. A. Casas, J. Garc\'{\i}a-Bellido \&
M. Quir\'{o}s [Nucl. Phys. {\bf B361}, 713 (1991)]
point out the problems one encounters
trying to put into agreement string cosmology with post-Newtonian
bounds at the present.

\newpage

\appendix
\section{Does $\Delta (\xi)$ vanish?}
In this Appendix we want to prove that the introduction of the
dilaton potential,
in Sect. IV, allows us to eliminate the
singularities in the fields and in the
energy density of the string bulk matter.
To show this, it is enough to study
the zeroes of $\Delta (\xi )$; we will
be able to avoid the singularities if
$\Delta (\xi ) \not= 0$ for $\xi$ real.
We have ($\xi_0 =0$ as in Sect. IV)
\begin{equation}
\Delta (\xi ) = \Omega - 2 \zeta \sqrt{\xi^2
+\xi_{1}^{2}} + \epsilon \xi^2
\label{AAdxi}
\,,\end{equation}
where we have defined
\begin{eqnarray}
\Omega &=& 4\beta - (1-\epsilon ) \xi_{1}^{2} \,,\nonumber
\\
\beta &=& \ell^2 V_0 - \case{1}{4} \sigma \,,\nonumber
\\
\sigma &=& 3 \alpha_{H}^{2} + n \alpha_{F}^{2} \nonumber
\\
\zeta &=& 3\alpha_H \hat\gamma + n \alpha_F \hat \lambda \,,\nonumber
\\
\epsilon &=& 1 - 3 \hat \gamma ^2 - n\hat\lambda ^2 \,.\nonumber
\end{eqnarray}
As we have yet stressed in Sect. III, we
must require $\epsilon >0$ to have a positive energy density.
It is immediately seen that $\Delta (\xi ) = 0$ implies
\begin{equation}
\epsilon^2 \xi^4 + 2 (\epsilon \Omega - 2 \zeta^2 ) \xi^2 + \Omega^2 -
4\zeta^2 \xi_{1}^{2} = 0 \,.
\label{AAxif}
\end{equation}
If we introduce
\begin{eqnarray}
y &=& \xi^2 \,,\nonumber
\\
a &=& \epsilon \Omega - 2 \zeta^2 \,,\nonumber
\\
b &=& \Omega^2 - 4\zeta^2 \xi_{1}^{2} \,,\nonumber
\end{eqnarray}
equation (\ref{AAxif}) reduces to
\begin{equation}
\epsilon^2 y^2 + 2 a y + b =0 \,.\nonumber
\end{equation}
Because $y=\xi^2$, $\Delta (\xi )$ does not
vanish in the real field if
and only
if one either of the following conditions is satisfied:
\begin{eqnarray}
& a) &\ \ \ \delta \equiv a^2 -\epsilon^2 b <0 \,,\nonumber
\\
& b) &\ \ \ -a+\sqrt{\delta}<0\ \ \ \ \ {\rm if}\ a>0,\ \delta >0 \,.
\nonumber
\end{eqnarray}
For what concerns condition $a)$ we have
\begin{eqnarray}
\delta &=& (\epsilon \Omega - 2 \zeta^2 )^2 - \epsilon^2
(\Omega^2 - 4
\zeta^2 \xi_{1}^{2} ) \,,\nonumber
\\
&=& -4 \zeta^2 (4\epsilon \ell^2 V_0 - \tau^2 ) \,,\nonumber
\end{eqnarray}
where
\begin{equation}
\tau^2 = \epsilon \sigma + \epsilon (1-\epsilon )\xi_{1}^{2}
+ \zeta^2 +
\epsilon^2 \xi_{1}^{2} >0
\label{AApi}
\end{equation}
(remember that $0<\epsilon <1$). Then $\delta <0$ implies
\begin{equation}
V_0 > {\tau^2 \over 4\epsilon \ell^2} >0\,.
\label{AAci}
\end{equation}
\par
Now with condition $b)$. We must take
\begin{equation}
\delta >0 \ \ \ {\rm and\ then}\ \ \ V_0 \le {\tau^2
\over 4\epsilon\ell^2}
\label{AAcii}
\,,\end{equation}
and
\begin{eqnarray}
a &>& 0 \,,\nonumber
\\
- a + \sqrt{\delta} &<& 0 \,.\nonumber
\end{eqnarray}
Let us start to study the condition $a>0$. It implies
\begin{equation}
\epsilon \Omega - 2 \zeta^2 = \epsilon [ 4 V_0 \ell^2
 - \sigma - (1 - \epsilon )
\xi_{1}^{2} ] - 2 \zeta^2 >0 \,,\nonumber
\end{equation}
and then
\begin{equation}
V_0 > {\eta^2 \over 4\epsilon\ell^2} >0\,,
\label{AAciii}
\end{equation}
where we have defined
\begin{equation}
\eta^2 = \epsilon \sigma + \epsilon (1-\epsilon ) \xi_{1}^{2} +
2 \zeta^2
= \tau^2 +\zeta^2 - \epsilon^2 \xi_{1}^{2} >0\,.
\label{AAeta}
\end{equation}
To make conditions (\ref{AAcii}) and (\ref{AAciii})
compatible we must
require $\eta^2 < \tau^2$, which means
\begin{equation}
\zeta^2 < \epsilon^2 \xi_{1}^{2}\,.
\label{AAciv}
\end{equation}
If inequality (\ref{AAciv}) is not verified, we have $\eta^2 > \tau^2$
and then the condition $a>0$ automatically implies $\delta <0$,
which means that $\Delta (\xi )$
never vanishes in the real field for condition $a)$.
\par
For what concerns the second condition, $\sqrt{\delta} < a$, we have
\begin{equation}
16 \epsilon^2 V_{0}^{2}\ell^4 - 8 \epsilon (\eta^2 -  2\zeta^2 )
 V_0 \ell^2 - 4
\zeta^2 \tau^2
+ \eta^4 >0 \,.\nonumber
\end{equation}
Because
\begin{equation}
V_0 = {\theta_\pm \over 4\epsilon\ell^2}\,,
\end{equation}
with $\theta_\pm = \eta^2 - 2\zeta^2 \pm 2 | \zeta
\epsilon \xi_1 |$, we must
require
\begin{equation}
4\epsilon V_0 \ell^2< \theta_- \ \ \ \ {\rm and}\ \ \ \ 4\epsilon V_0
\ell^2 > \theta_+\,.
\label{AAcv}
\end{equation}
Summing up, condition $b)$ is equivalent to the following
conditions on $V_0$
\begin{mathletters}
\label{AAcvi}
\begin{eqnarray}
\eta^2 < & 4\epsilon V_0 \ell^2 & \le \tau^2\,,
\label{AAcvi a}
\\
\theta_+ < & 4 \epsilon V_0 \ell^2 &\,,
\label{AAcvi b}
\\
& 4 \epsilon V_0 \ell^2 & < \theta_-\,.
\label{AAcvi c}
\end{eqnarray}
\end{mathletters}
Because $\zeta^2 < \epsilon^2 \xi_{1}^{2}$ we have also
\begin{eqnarray}
\eta^2 < & \theta_+ & < \tau^2 \,,\nonumber
\\
& \theta_- & < \eta^2 \,,\nonumber
\end{eqnarray}
and then, conditions (\ref{AAcvi}) reduces to
\begin{equation}
0 < {\theta_+ \over 4\epsilon\ell^2} < V_0 \le
{\tau^2 \over 4\epsilon\ell^2}
\,.\label{AAcvii}
\end{equation}
In any case, the request that $\Delta (\xi )$ never
vanishes (conditions $a)$ and
$b)$) automatically implies $V_0 >0$ (inequalities (\ref{AAci}) and
(\ref{AAcvii})).
\section{Representation in the Einstein Frame.}
It is well known that scalar-tensor and non-linear
gravity theories can be
reformulated in a more conventional framework:
Einstein gravity plus a
minimally coupled scalar field \cite{BCMS}.
We have only to perform a Weyl rescaling of the
metric tensor, $g_{\mu\nu} \to \Omega^2 g_{\mu\nu}$. For ST we must
take \cite{LESEA}
\begin{equation}
\Omega^2 = e^{2\Phi /(D-1)} \label{WT}
\,.\end{equation}
Then the action (\ref{lesea}) reduces to
\begin{equation}
S \to - {1 \over 2\kappa^2} \int d^D x \sqrt{-g} \left\{ R - {1 \over D-2}
(\partial \Phi )^2 + c e^{-{D-1 \over D-2} \Phi}\right\}
\,.\label{wtlesea}
\end{equation}
\par
Although the BD frame,\footnote{The representation in which the
scalar field couples non-minimally to the scalar curvature, action
(\ref{lesea})} seems the most natural in ST, it is also
interesting to study what happens
to solutions (\ref{unsol}) when we perform the Weyl rescaling
(\ref{WT})\footnote{For a review on the debate about the
two frames see \cite{MS}}.
\par
Recently it was stressed \cite{IDFI,DPSC} that the
`pre-big-bang' era of the BD
frame is naturally mapped to an accelerated contraction phase
in the Einstein frame.
But this is not true in general. The situation
is more complex and needs a more
accurate analysis. We will deal with non-singular solutions,
although for the asymptotic behavior the same results can be
applied to solutions (\ref{gss}) as well.
\par
Because we have not an analytical expression for
the scale factor, we can
extract qualitative informations looking at
the dynamics of the transformed
Hubble parameter\footnote{We could have
had informations about the correct
(qualitative and quantitative) dynamics
of the scale factors by numerically
integrating the equations. But, because
there are many free parameters which
strongly determine the evolution, it
is more interesting to study the shape of
the Hubble parameter.}.
\par
After performing the Weyl transformation (\ref{WT}) we get
(a prime denotes differentiation with respect to $\xi$)
\begin{eqnarray}
{a' \over a} &\to & \left( {a' \over a}\right)_{{\rm E}} =
{e^{\Phi /(n+2)}
\over n+2} \left( - \bar\Phi ' + (n-1) {a' \over a} -
n {b' \over b} \right)
\\
{b' \over b} &\to & \left( {b' \over b}\right)_{{\rm E}} =
{e^{\Phi /(n+2)}
\over n+2} \left( - \bar\Phi ' -3 {a' \over a} +2 {b' \over b}
\right)
\,,\end{eqnarray}
and then, substituting the equations (\ref{equat}) \cite{TESI},
\begin{mathletters}
\label{HWT}
\begin{eqnarray}
\left( {a' \over a} \right)_{{\rm E}} &=& {2 e^{\Phi
/(n+2)} \over (n+2)\Delta
(\xi )} \left[ (n -1) \alpha_H - n\alpha_F  + \xi +
((n-1)\hat\gamma -
n\hat\lambda ) \sqrt{\xi^2 + \xi_{1}^{2}} \right]
\,,\label{HWT a}
\\
\left( {b' \over b} \right)_{{\rm E}} &=& {2
e^{\Phi /(n+2)} \over (n+2)\Delta
(\xi )} \left[  2 \alpha_F  -3\alpha_H + \xi +
(2\hat\lambda -3\hat\gamma )
\sqrt{\xi^2 + \xi_{1}^{2}} \right]
\,.\label{HWT b}
\end{eqnarray}
\end{mathletters}
Let us introduce
\begin{eqnarray}
\{ A_i \} &=& \left( (n-1)\alpha_H - n \alpha_F\, ,\,
-3\alpha_H + 2 \alpha_F \right)\,,
\\
\{ B_i \} &=& \left( (n-1) \hat\gamma -
n\hat\lambda\, ,\, - 3\hat\gamma + 2 \hat\lambda
\right) \,.
\end{eqnarray}
Then the asymptotic behavior of the scale factor
depends on the sign of
$\xi/|\xi | + B_i$: for $\xi/|\xi | + B_i >0$
we have expansion,
$\xi/|\xi | + B_i <0$ means contraction,
while $\xi/|\xi |
+ B_i =0$ correspond to a stabilization
of the scale factor. The
Hubble parameter changes sign in the interval
\begin{equation}
{- A_i - \sqrt{ A_{i}^{2} + (B_{i}^{2} -1)(A_{i}^{2} -
B_{i}^{2} \xi_{1}^{2})}
\over 1 - B_{i}^{2}} < \xi <
{- A_i + \sqrt{ A_{i}^{2} + (B_{i}^{2} -1)(A_{i}^{2} -
B_{i}^{2} \xi_{1}^{2})}
\over 1 - B_{i}^{2}}\,,
\end{equation}
if $A_{i}^{2} + (B_{i}^{2} -1)(A_{i}^{2} - B_{i}^{2} \xi_{1}^{2}) >0$.
Then it
can happen that we start in the string frame with a
scale factor which presents
a transition expansion-contraction-expansion,
but in the Einstein frame it
experiences a monotonic expansion. It is then
evident that it is not generally
true that the expanding pre-big-bang phase
is always converted to a
contraction by Weyl rescaling, as stated
in \cite{IDFI} . It happens only for
homogeneous and isotropic models, and
when $B_i -1 <0$ with $\omega_i >0$ (we
remember that
$\{ \omega_i \} = (\hat\gamma , \hat\lambda )$).
It is worth to notice
that after the Weyl rescaling (\ref{WT}), we have
 eliminated the
non-minimal coupling of the scalar curvature with
the dilaton, but we have
still a theory with dynamical gravitational
coupling, because of the presence of
the volume of the internal space. So, if we
want that the two representations
be in accordance with observational
constraints about the variability of the
4-dimensional Newton's constant,
we must impose the two conditions $\hat\gamma
=\case{1}{3}$ and $\hat\lambda =0$.

\newpage
\begin{figure}
\caption{The evolution of the scale factor for
($a$) $\alpha_i /\omega_i >0$ and ($b$) $\alpha_i /\omega_i <0$.
In the second case, for $\xi >\xi_+$ the scale factor
starts from infinity, contracts to a minimal value and then expands
monotonically. Even if $a(\xi_+ )\not= 0$ the curvature has in $\xi_+$ a
singularity.}
\label{fig1}
\end{figure}
\begin{figure}
\caption{The evolution of
the scale factor, as result of numerical integration,
 for ${\rm sign} [\alpha_i \omega_i ] =-1$ for various values of
$\alpha_i$. The scenario describes a background with a
non-monotonic dynamics.
The universe starts expanding for large negative time; then for $|\xi|<
\left[(\alpha_i /\omega_i )^2 -\xi_{1}^{2}\right]^{1/2}$ a
contraction phase is present before the universe restart
expanding.}
\label{fig2}
\end{figure}
\begin{figure}
\caption{Evolution of
the Hubble parameter $H_i$ for
${\rm sign} [\alpha_i \omega_i ]=-1$. The bold lines
represent the phases of
accelerated evolution during which the problems of the SBB
(flatness, horizon, structure formation, $\ldots$)
can be solved.}
\label{fig3}
\end{figure}
\begin{figure}
\caption{4-dimensional ``Newton's constant''
($G_N \sim b^{-n}e^{\Phi}$)
(solid line) and internal scale factor, $b(t)$ (dashed lines) as
a function of the coordinate time $\xi$. Asymptotically both $G_N$
and $b(t)$ tend towards a
constant value in accord with observational constraints,
while experiencing a
non-trivial dynamics for short times.}
\label{fig4}
\end{figure}
\end{document}